# Near-field optical imaging with a CdSe single nanocrystal-based active tip

**Y. Sonnefraud, N. Chevalier\*, J.-F. Motte, S. Huant**

*Laboratoire de Spectrométrie Physique, Centre National de la Recherche Scientifique and Université Joseph Fourier - Grenoble, BP 87, 38402 St Martin d'Hères, France*
*\*Present address : Commissariat à l'Energie Atomique, Laboratoire d'Electronique et des Technologies de l'Information, 17 rue des Martyrs, 38054 Grenoble Cedex 9, France*
*serge.huant@grenoble.cnrs.fr*

**P. Reiss, J. Bleuse, F. Chandezon**

*Commissariat à l'Energie Atomique, Département de la Recherche Fondamentale sur la Matière Condensée, 17 rue des Martyrs, 38054 Grenoble Cedex 9, France*

**M. T. Burnett, W. Ding, S. A. Maier**

*Centre for Photonics and Photonic Materials, Department of Physics, University of Bath, Bath, BA2 7AY, United Kingdom*

**Abstract:** We report near-field scanning optical imaging with an active tip made of a single fluorescent CdSe nanocrystal attached at the apex of an optical tip. Although the images are acquired only partially because of the random blinking of the semiconductor particle, our work validates the use of such tips in ultra-high spatial resolution optical microscopy.

## 1. Introduction

Near-Field Scanning Optical Microscopy (NSOM) is a scanning probe technique that allows overcoming the diffraction limit in optics [1]. In NSOM, the lateral resolution is basically limited by the size of the light source – the tip - as well as by the tip-surface distance which can be brought to arbitrary small values by suitable feedback schemes [2]. In its so-called aperture configuration, NSOM uses a tapered and metal coated optical fiber that can offer resolutions down to 30 nm [3]. However, such resolutions are reached in specific cases only, e.g., in spectroscopic studies at low temperature where quantum emitters produce extremely sharp lines that are clearly distinguished from an unstructured background. A more realistic resolution limit is at 50 nm. This is mainly due to the difficulty of controlling the tip fabrication for small apertures as well as to the penetration of light into the tip metal coating.

Routes to push the resolution in aperture NSOM down to the nanometer range have been opened up by applying the concept of active tips. Such tips are made of a NSOM probe on which a fluorescent nano-object has been attached. The fluorescence emitted by the object is used as source of light whose size matches roughly with that of the active object. A spectacular validation of this concept has been reported by Michaelis et al [4]. These authors used a single terrylene molecule – in essence the smallest source of light that can be thought of - embedded in a p-terphenyl microcrystal. They reported a resolution of 180 nm at low temperature. This resolution was limited by the difficulty of selecting a properly located single molecule in a rather large microcrystal. Other types of active tips have been investigated, such as for example nitrogen-vacancy color centers in diamond [5], rare-earth doped glass particles [6] and gold nanoparticles. In the latter case one does not use the particle fluorescence, but its field enhancement or the detection of substrate-induced shifts in the plasmon resonance of the particle [7-9].

Semiconductor nanocrystals (NCs) have also been used because they offer promising properties such as natural nanometer-scale sizes and room-temperature operation. Shubeita *et al.* [10] demonstrated a resolution of 170 nm in a fluorescence resonance energy transfer experiment using CdSe NCs as excitation donors. The limited resolution in this experiment is due to the large number of NCs involved in the optical signal. Very recently, Aigouy et al. performed NSOM fluorescence imaging with a 200nm silica sphere covered with CdSe NCs [11]. Quite good optical contrast was achieved though without any improvement in spatial resolution because of the large size of the fluorescent probe. Recently, Chevalier et al [12] have realized active tips containing only a few active CdSe NCs at the apex, actually down to a single NC. However, no optical image has been reported with such active tips so far. The aim of this paper to present and comment on the very first images obtained with such CdSe-NC based active tips.

## 2. Experiment

*2.1 Microscope setup*

The optical experiments discussed in this paper have been conducted on a home-made fluorescence microscope that can operate both in confocal or NSOM modes. This setup has been developed on an inverted microscope. A schematic drawing is given in Fig. 1. In the confocal mode, both the excitation at 458 nm wavelength and the light collection are ensured by a ×60, NA 0.95 dry objective. The sample is placed in a holder driven by a piezoscanner. Before reaching the detector, the collected light is filtered by a dichroic mirror at 458nm that cuts the excitation light. A bandpass filter (542-622 nm) centered on the emission line of the CdSe NCs further narrows the transmission window. In NSOM, which operates in transmission, the tip is glued to a quartz tuning fork [2]. Tip-sample distance is kept constant with a feedback loop acting on the vibration amplitude of the tuning fork mechanically driven at resonance. One should note that in our setup, the only moving part is the sample. As the sample surface is placed in the focal plane of the collection objective which has a field depth of 300 nm, the sample never exits the focus volume during operation. Therefore, the optical signal is not harmed in such a configuration. For the detection, we either use a spectrometer equipped with a charge-coupled device (CCD) camera cooled at 110 K for spectral studies, or an avalanche photodiode (APD) in the photon counting mode for imaging. The experiments have all been conducted at room temperature and in the air.

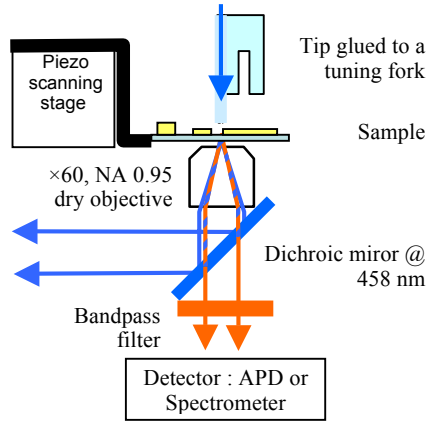

Fig. 1. Scheme of the home-made confocal and NSOM microscope.

*2.2 CdSe nanocrystals*

Our NCs have been synthesized following the procedure presented in [13]. They are CdSe-ZnSe core–shell NCs and have been functionalized with tridecaneditioic acid [14] and diluted in toluene. They display very narrow size dispersion around a diameter of 3.8 nm and a fluorescence peak centered at 582 nm for the ensemble. At the single particle level, they exhibit a distinctive blinking behavior, i.e., they randomly switch between emitting "on" and non-emitting "off" states [15]. This phenomenon is clearly seen in Fig. 2 (a) which shows a confocal fluorescence image of NCs that have been spin coated on a glass cover slip. The "off" periods can be seen as horizontal dark stripes in some fluorescence spots. As a consequence of this blinking, only partial images are acquired. Figure 2 (b) shows a spectrum of the isolated object encircled in grey in Fig. 2 (a) together with a spectrum of the ensemble for comparison. The two spectra have only slightly different widths. However, spectra acquired for shorter integration times (15s) on the same single object, see Fig. 2 (c), have definitively smaller full width at half-maximum (FWHM), around 27 nm, compared with 32 nm for the ensemble. This smaller FWHM for single objects agrees with previous observations and is a fingerprint for a single particle behavior. [12]

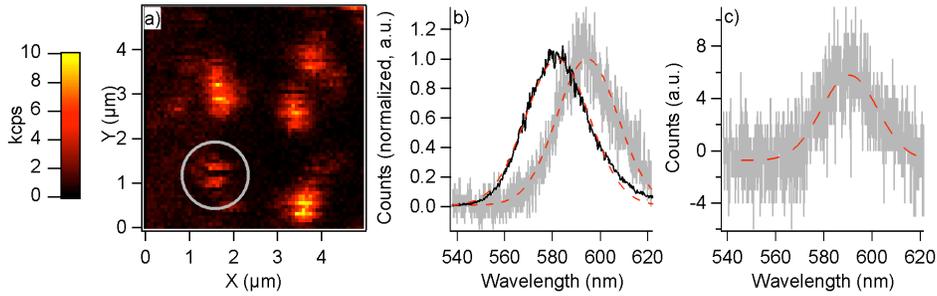

Fig. 2. Blinking behavior of single CdSe NCs. (a) A confocal image of NCs deposited on glass. The scanning area is 5×5μm². The excitation density is 200 W.cm$^{-2}$ and the integration time of the APD is 10 ms per pixel. The color scale is in kilo counts per second (kcps). (b) Normalized spectra of a NC ensemble (black) and of the single object (grey) encircled in grey in Fig. 2(a). The integration time of the single particle spectrum is 60s. Red dashed lines give Gaussian fits to the spectra with a center wavelength at 582 (595) nm and a FWHM of 31.7 (30.2) nm for the ensemble (single NC). (c) The spectrum of the single NC integrated over 15s has a FWHM reduced to 27.4 nm.

*2.3 Test samples for imaging*

The test samples used for NSOM imaging are 40 nm thick gold structures deposited by electron beam lithography on glass cover slips. Prior to gold deposition, a 5 nm-thick layer of NiCr is coated on the glass substrate to offset charging and improve gold sticking. Figure 3 shows a SEM image of a typical test sample compared with its NSOM transmission image obtained at 458 nm wavelength. Here, we used a regular, i.e., not stained with NCs, optical tip without any optical filtering. In both images, details in the test structure can be seen with quite good contrasts and resolutions. In the NSOM image for example, the standard deviation in the signal is about 11%. This is possibly due to remaining instabilities in the light injection into the optical fiber. Line cuts through the thin triangle summit or the smaller 500 nm in-diameter disk entail a spatial resolution of 180 nm, in good agreement with the aperture size of 200 nm. This imaging with a regular tip confirms the good operation of our NSOM; this was important to check for the purpose of the active tip experiments described below.

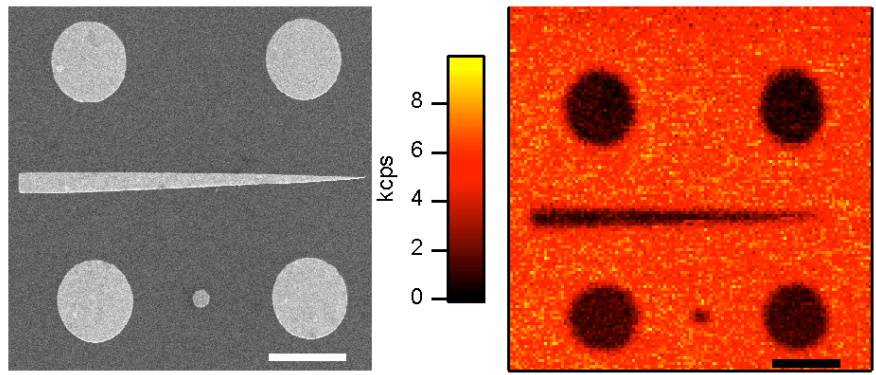

Fig. 3. Test samples for NSOM imaging. [left] SEM image of the sample. Scale bar: 2μm. Bright parts are the gold structures and the dark background is the glass substrate. [right] Corresponding NSOM straight transmission image at 458 nm wavelength. Note the inverted contrast due to light reflection on the gold patterns. Scale bar: 2μm. The color scale refers to the optical image.

*2.4 Active tips*

The active tips using these NCs as active medium are produced in a two-step process. We first start from aluminum-coated optical tips that are prepared following the modified chemical-etching method described in [16]. Special care was taken to select a good substrate tip according to the quality criteria recalled in [16], namely a reasonable transmission efficiency and a smooth far-field emission profile that allows inferring an aperture size. In the example described below, the initial optical tip has an aperture of 200 nm and a transmission of $10^{-2}$.

The second step consists in dipping the initial probe into a low-mass PMMA solution stained with a low-concentration solution of NCs; see [12] for an extensive report on this method. Subsequent characterizations including fluorescence spectra and/or time-traces recording of the emission rate ensure that a limited number of nano-objects are active at the tip apex; see [12] and the example below.

The active tip is then inserted in the NSOM head and is driven to feedback over the sample. The tip is first maintained above the transparent part of the sample close to the gold structure. Spectra are acquired to confirm the presence of emitting NCs at the apex. Two successive spectra are shown in Fig. 4. They both have a FWHM slightly thinner than the ensemble indicating that a limited number of NCs, possibly only one, contribute. These spectra are superimposed on a background due to the parasitic fluorescence occurring in the optical fiber guide. [12]

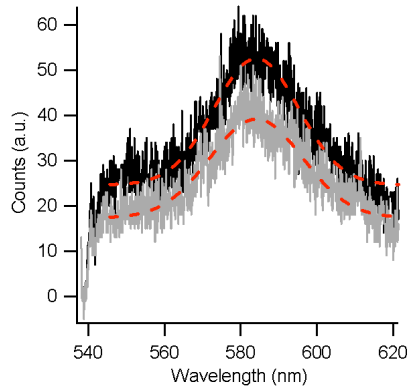

Fig. 4. Successive spectra of the active tip. The upper (lower) spectrum is the first (second) acquisition. The integration time is 60s and the excitation density is 800 W.cm$^{-2}$. Dashed red lines are Gaussian fits to the peaks with a center wavelength at 584 nm for both spectra and a FWHM of 26.2 nm (28.9 nm) for the upper (lower) spectrum.

**3. Results and discussion**

Figures 5 (a)-(b) show two NSOM images of a test sample acquired with the very tip described above. Here, the bandpass filter is used to restrict the signal to the spectral range of the NC emission. The image of Fig. 5(a) has been acquired while the tip scanned from left to right whereas the image of Fig. 5(b) has been acquired during the backward scan. Therefore, the two images have been recorded almost simultaneously with a time shift of at most 6.6 s for each line, i.e., the time required to complete a horizontal back and forth line scan. The main parts of both images display a signal level around 1.5 kcps that falls down over the gold structures (dark zones). Interestingly enough, a few lines show an intermittently higher signal, up to 3 kcps (brighter pixels). This is especially marked between the two upper disks.

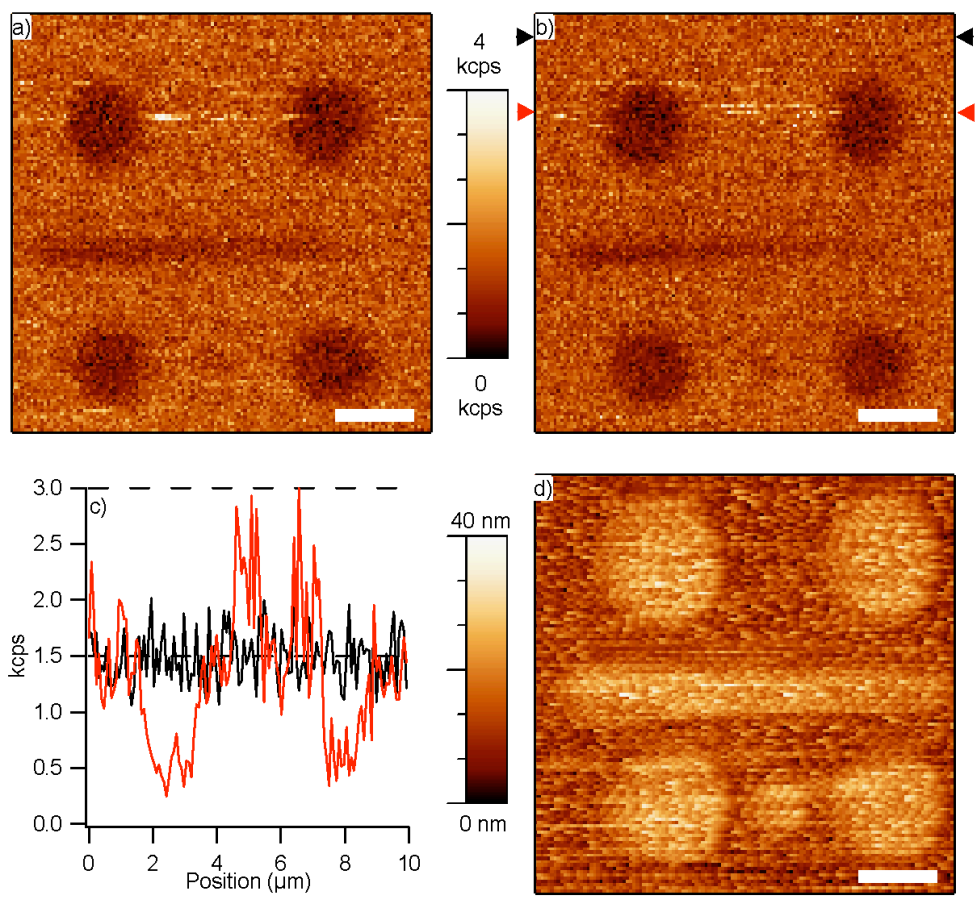

Fig. 5. Imaging with an active tip. (a) Forward scanning image. Light is filtered in the spectral domain of the CdSe NC emission. Image sizes are 128 pixel ×128 pixel and 10 μm ×10 μm. The scale bar is 2μm and the scan speed is 5 μm.s$^{-1}$. The integration time of the APD detector is 10 ms per pixel. The excitation density is around 1.6 kW.cm$^{-2}$. (b) Same as (a) for backward scanning. (c) Cross cuts through the image in Fig. 5(b). Red (black) is a cross cut along a line presenting an increase in intensity (a background line), whose positions are indicated in Fig. 5(b) by the red (black) arrows. (d) Topography image acquired simultaneously with Fig. 5(b).

To better visualize these bright pixels, Fig. 5(c) shows two cross cuts through the image of Fig. 5(b). The first cross cut is in a zone above the transparent part of the sample which does not exhibit such bright pixels. One can see a signal fluctuating around 1.5 kcps due to background fluorescence. The standard deviation here is 16% which is slightly higher than in standard NSOM (cf. Fig. 3). The second cross cut is along a line with bright pixels. It shows that the signal rises far above the standard noise in the first cross cut. Indeed, the standard deviation rises above 30% in this region of interest.

We argue that the bright pixels of Figs. 5(a)-(b) form the useful signal due to the active tip. First, the main source of light which allows revealing a contrast between the opaque gold structures and the glass substrate is the fiber fluorescence discussed above. This background is seen in the tip spectra of Fig. 4 as well. Its intensity is close to 1.5 kcps on the optical image. Second, several clues lead us to claim that the intermittent doubling [12] of the optical signal level in Figs. 5(a)-(b) is due to the fluorescence of a single blinking NC attached to the tip apex. Indeed, the spectra of the tip taken before operation clearly confirm the presence of (a) NC(s) at the tip apex as already mentioned, see Fig. 4. The FWHM is smaller than in the ensemble and coincides with that of a single NC measured in confocal microscopy; see Fig. 3(b). Moreover the topographic image taken simultaneously with the optical image of Fig. 5(b) is shown on Fig. 5(d). While this image is further limited in resolution as compared to the optics because of the metal coating of the substrate tip, it does not show any artifact in the region of high optical signal. This confirms the optical origin of the bright pixels.

It should be mentioned that all images recorded after those of Fig. 5 only displayed the background fiber signal and did not show any intermittent increase of the signal. This very fact could be a sign of extremely long "off" times [12, 15] or of photo-bleaching of the NC. Photo-bleaching in such NCs is usually attributed to the oxidation of the surface. [17] This is likely to occur in our experiments performed in the air.

As far as the anticipated improved resolution is concerned, the acquired images are too partial to draw any definitive conclusion. However, it is worth noting from the bright lines in Figs. 5(a)-(b) that some bright pixels seem to "enter" into the dark zones. This is more apparent on the dark disk in the upper right corner of the optical images. Indeed, concentrating on the cross cut of Fig 5(c), the edge of the gold dot on the right seems to be sharper when the signal has increased because of the NC emission (position around 7 µm) than when there is only background signal (position around 9 µm). This goes in the direction of an improved resolution with the single NC based active tip, compared with the background image. Indeed, the latter is limited in resolution by the larger size of the substrate tip. To confirm this trend, it will be necessary to reduce the impact of photo-bleaching by conducting future experiments under neutral, e.g. argon, atmosphere. Furthermore, CdSe/ZnSe/ZnS core/shell/shell nanocrystals could be used, which exhibit enhanced photo-stability due to the additional protective ZnS outer layer. Finally, it should be possible to improve the NC attachment process by using a "select and fishing" method [8] combined with a FIB milling of the substrate tip. [18]

## 4. Conclusions

We have presented the very first NSOM images taken with an active optical tip made of a single CdSe NC attached at the apex of a regular tip. This tip operates under ambient condition and has the potential for an ultra high spatial resolution in optics. Although the images recorded so far are too partial to infer a spatial resolution because of the NC blinking, they demonstrate that imaging with a single NC as light source is possible.

**Acknowledgments**
Helpful discussions with J. Plain, G. Lérondel and P. Royer are gratefully acknowledged. We thank support from the "Action Concertée Nanosciences 2004" (NANOPTIP project). This work was in part sponsored by the Engineering and Physical Science Research Council (EPSRC UK).